# A tilted pulse-front setup for femtosecond extreme ultraviolet transient grating spectroscopy in highly non-collinear geometries

A. Battistoni[1,2], H. A. Dürr[3], M. Gühr[1,4], T. J. A. Wolf[1,*]

[1]Stanford PULSE Institute, SLAC National Accelerator Laboratory, Menlo Park, CA 94025, USA

[2]Department of Physics, Stanford University, Stanford, CA 94305, USA

[3]SIMES, SLAC National Accelerator Laboratory, 2575 Sand Hill Rd., Menlo Park, CA 94025, USA

[4]Institut für Physik und Astronomie, Universität Potsdam, 14476 Potsdam, Germany

*thomas.wolf@stanford.edu

**Abstract:** We demonstrate a tilted pulse-front transient grating technique that allows to optimally utilize time-resolution as well as transient grating line density while probing under grazing incidence as typically done in extreme ultraviolet (EUV) or soft x-ray (SXR) experiments. Our optical setup adapts the pulse front tilt of the two pulses that create the transient grating to the relative tilt grazing incident pulse. We demonstrate the technique using all 800 nm femtosecond laser pulses for transient grating generation on a vanadium dioxide film. We probe that grating via diffraction of a third 800 nm pulse. The time resolution of 100 fs is an improvement by a factor 30 compared to our previous experiments on the same system (1,2). The scheme paves the way for EUV and SXR probing of optically induced transient gratings on any material.

**Introduction**

Transient grating (TG) spectroscopy is a particular form of four wave mixing (FWM) spectroscopy providing time resolved information about the samples third order susceptibility. The grating is formed by two identical laser pulses that hit the sample simultaneously under a certain angle, which translates into a spatial modulation of the excitation. A third pulse is scattered from that grating with variable delay after the excitation. Since the beginning TG spectroscopy has been used on numerous samples (3–5) to elucidate the complex dynamics on molecules and solid materials. An implementation of TG spectroscopies in the Extreme Ultraviolet (EUV) and Soft X-Ray (SXR) regimes allows new insights in the dynamics of materials by expanding the possible excitation energy range, enhancing the spatial resolution to sub-molecular length-scales and enabling elemental selectivity through core resonances. In particular, EUV-SXR FWM can have widespread applications in studying charge/energy transfer phenomena, exciton dynamics in strongly correlated systems, and quasi-particle properties in thin films and surfaces.

The light sources are readily available to the steep progress in Free Electron lasers (6–9) and high harmonic generation, recently even progressing from the EUV towards the SXR domain (10–13). Experiments with EUV excitation pulses became possible at the FERMI free electron laser reaching unprecedented grating line densities (14–16). The properties of HHG radiation like full temporal and spatial coherence and broad coherent bandwidth make it attractive for FWM spectroscopy. So far, HHG sources have been used as probes for optically induced transient gratings, allowing to study a whole plethora of known optically induced excitations using an element sensitive probe. Those experiments optimally use the intrinsically low photon flux of HHG sources and first demonstrations showed element sensitivity at metal M-edges as well as superb sensitivity on small surface changes (1,2,17–19).

To fully harvest the potential of ultrafast optical pump - short wavelength probe TG spectroscopy to investigate surfaces and 2D materials, two points need to be considered: light in the EUV/SXR domain is only efficiently reflected at grazing incidence[1] and a high line density grating is needed to efficiently diffract short wavelengths. To produce highest possible line density on the sample, the optical excitation pulses best hit the sample under normal incidence. Due to the pulse front mismatch between pumps and probe however, the time resolution is compromised, which poses a major obstacle to investigate processes on the sub-ps timescale. In a past experiment studying the insulator-to-metal transition of vanadium dioxide with the TG technique, we succeeded in pinning down the phase transition at the vanadium M-edge. However, the time resolution was smeared out to 3 ps due to the pulse front angle mismatch (1,2).

In this work, we present a modified design for the TG setup. We correct the pulse front mismatch by tilting the pulse-front of the pump beams. We demonstrate a time resolution better than 100 fs, paving the way for future ultrafast EUV-SXR probe TG experiments. We use, as a prototypical sample, the well-studied $VO_2$ films, a strongly correlated system well known for its insulator to metal transition (IMT) that can be induced by slight temperature changes or by light excitation (20–25).

**Experimental Setup**

The experimental setup is a modification of a HHG TG setup described in detail elsewhere (1). In short, we are using the output of a commercial Ti:Sapphire laser system (30 fs, 800 nm), which is split into a pump and a probe beam path. The setup is equipped with a high vacuum HHG beamline in the probe beam path (26), which was however not used as such for the present demonstration experiments. By removing an aluminum filter and the HHG gas, the 800 nm HHG seed beam can be transmitted and focused by the EUV optics onto the detector plane. It intersects with the sample before the focus in a grazing angle of $\theta'_{pr} = 22°$[2]. Transient gratings were excited on the sample surface by two pump pulses with incidence angles of $\theta_{pu} \approx 2°$ (See Fig. 1).

The angle $\theta'_{pr}$ is not strongly grazing, the modifications to the original setup mainly concern the generation of the two grating-exciting pulses in the pump beam path. Due to the finite size of the pump and probe beams on the sample, the large difference in angle of incidence leads to a considerable pulse front mismatch. Therefore, the achievable time resolution is on the order of 3 picoseconds despite the approx. 30 fs duration of all involved light pulses.

---

[1] Unless the material under investigation is grown as a Bragg stack facilitating high normal incidence reflection
[2] This angle is not strongly grazing, however, its effect on the time resolution is already catastrophic. For shorter wavelength, even shallower angles need to be used to obtain good reflectivity from the sample.

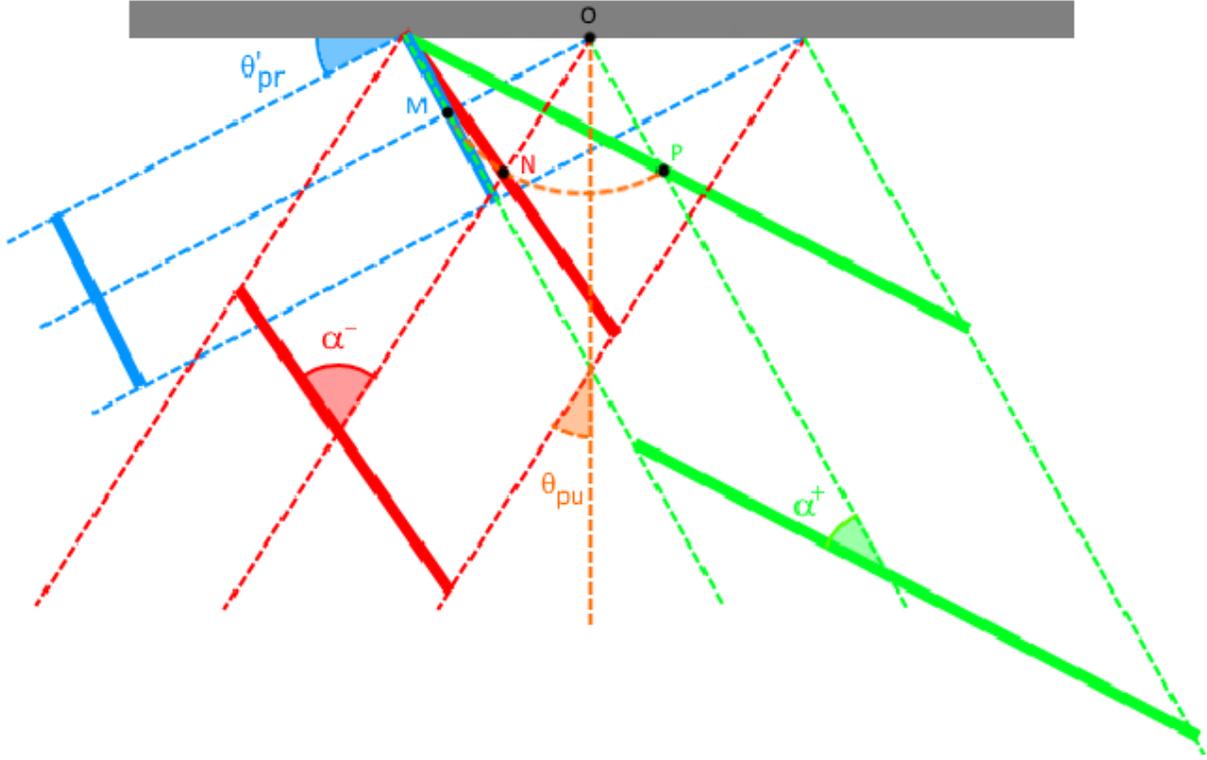

*Figure 1: Geometry of pump and probe beam paths at the sample position: the two pump beams intersect with the sample in an angle of $\theta_{pu} \approx 2°$ with respect to the surface normal, whereas the probe beam is irradiated on a mild but, for the time resolution, not negligible grazing angle ($\theta'_{pr} = 22°$). For optimum temporal overlap with the probe pulses over the whole illuminated area, the pulse fronts (thick lines in the sketch) of the two pump beams have to be tilted to match the probe pulse front. The angles in the figure are overstated for better visibility.*

In order to improve the time resolution into the <100 fs domain, the pulse-fronts of pump and probe pulses need to be matched. This can be achieved by either tilting the fronts of the pump beams or the probe beam. In principle, it would be more straight forward to tilt the pulse front of the probe pulse since this only affects one beam. However, this is challenging in the case of SXR pulses. We therefore employed a scheme to tilt and match the pulse fronts of the optical pump pulses. The pulse front tilt angles ($\alpha^+$ and $\alpha^-$) for the two pump pulses have to satisfy the following non-trivial expression:

$$\alpha^\pm = arcsin\left(\frac{\cos\theta_{pu}}{B^\pm}\right) \text{ with } B^\pm = \left[\left(\cos\theta'_{pr} \pm \sin\theta_{pu}\right)^2 + \cos^2\theta_{pu}\right]^{1/2} \quad (1)$$

It is possible to obtain such expression imposing that the three beams, overlapping each other on the sample, uniformly keep the pump-probe time delay constant. This condition, referring to Fig. 2, is satisfied when $\overline{MO} = \overline{NO} = \overline{PO}$.

The matching condition could be fulfilled by individually tilting the fronts of the two pump pulses. Aiming to reduce the degrees of freedom in the alignment procedure, we employ a more elegant design (27) for the production and recombination of the two pump beams, which intrinsically matches their pulse fronts: we employ a combination of a blazed grating (BG in Fig. 2), which generates the pulse front tilt, with a transmission grating (``phase mask'' - PM in Fig. 2 - *Holo/or*, DS-234-800-Y-A, optimized for λ=800nm, fused silica substrate, 22.5µm line period), which separates the two pump pulse arms in its ±1 diffraction orders. Pulse front matching is ensured by confocal imaging the transmission grating on the sample surface (S) (28).

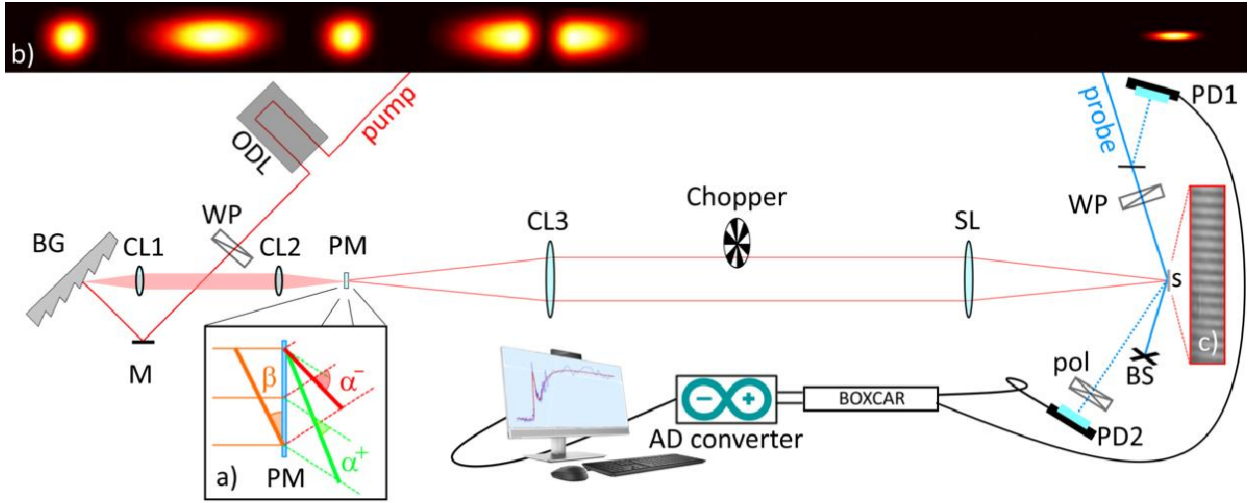

*Figure 2: Optical setup generating a transient grating with pulse fronts matched to the probe pulses on the sample surface: the pump beam is sent, after an optical delay line (ODL), to a tilting pulse-front blazed grating (BG) via the folding mirror M. The first diffraction order is imaged horizontally, by two confocal cylindrical lenses CL1 and CL2, on the PM. The latter is optimized to split the beam in the ±1 diffraction orders. The two orders are recombined by simultaneously imaging the PM in horizontal direction and focusing the beams in vertical direction on the sample (S) through the combination of the cylindrical lens CL3 and the spherical lens SL. The probe beam reaches the sample at grazing incidence and, interacting with the TG, generates the FWM signal which is then detected in 1st order direction. The 0th order is blocked by a beam stop (BS). The pump beam polarization can be tuned by a wave-plate (WP). Fig. 2-a shows the pulse-fronts of the beams at the PM. Fig. 2-b: a sketch of the pump beam profile plotted directly on top of the corresponding positions along the beam-path. Fig. 2-c: the actual pumps interference pattern measured with a CCD camera with 5:1 imaging magnification.*

Combining the grating equation and eq. 1, it can be shown that the ±1 diffraction orders after the PM have the desired pulse-front tilt if the PM is orthogonal to the 800 nm beam direction and the latter has a pulse-front tilt β (see Fig. 2-a) so that

$$\tan \beta = \left(\cos \theta'_{pr}\right)^{-1} \tag{2}$$

Comparing eq. 1 and eq. 2, the simplification introduced with this design is self-evident. We would like to stress the fact that β is only a function of the probe incident angle $\theta'_{pr}$ not the incidence angles of the pump pulses. This allows for modifying the line spacing of the TG independently of the pulse front tilt by replacing the PM.

To induce a pulse front tilt β at the PM position (Fig. 2-a), we use a reflective diffraction grating with the characteristic equation

$$\sin \gamma + \sin \delta = Nm\lambda \tag{3}$$

where γ, δ, N, m, and λ are the angle between the incoming beam and the normal to the grating surface, the diffraction angle, the grooves density of the grating, the diffraction order and the wavelength of the scattered light. Here, we use the convention δ if at the same side of γ in respect to the normal to the grating surface. It is easy to show that the 1st diffraction order is characterized by a pulse-front tilt β so that

$$\sin\gamma = \frac{\cos(\delta + \beta)}{\sin\beta} \tag{4}$$

Using a reflecting blazed grating (BG in Fig. 3, Thorlabs, N=830 lines/mm), we can satisfy both eq. 2 and eq. 4 with $\delta = 44.26°$ and $\gamma = -1.94°$. In order to compensate for the horizontal angular dispersion due to the diffraction, we use two confocal cylindrical lenses (CL1 and CL2 in Fig. 3, both f = 5 cm) to image the grating on the PM. Imaging the BG surface onto the PM instead of focusing the pump beam into the PM has the additional advantage that intensities are kept well below thresholds for nonlinear effects and damage in the PM. By employing the combination of a cylindrical lens (CL3) and a spherical lens (SL, both f = 15 cm), we simultaneously image the PM in the horizontal plane onto the sample surface and focus the pump beams in the vertical plane. The resulting spot shapes are sketched in Fig. 3-b for several positions in the beam path. By using the horizontal imaging / vertical focusing scheme, we ensure efficiently pumping the complete sample area, which is irradiated in grazing incidence by the probe pulse. This area is additionally increased by the fact that the probe pulse intersects with the sample before its focus in the detector plane. The optical setup allows to conveniently alternate the polarizations of pump and probe beams using halfwave plates.

To demonstrate the functionality of the optical setup, we performed all 800 nm transient grating experiments in vertical polarization at room temperature on a 100 nm thick single-crystal $VO_2$ film, which was prepared by pulsed laser deposition on a $Al_2O_3$ ($10\bar{1}0$) substrate (29). The employed fluences were well below the damage threshold (< 20 mJ/cm$^2$ (1)). The probe beam had an estimated fluence of 1 mJ/cm$^2$. The delay between excitation and probe pulses was implemented via a computer controlled mechanical delay stage. The transient grating signal was detected by a photodiode in first-order diffraction direction of the transient grating. To avoid contamination of the transient signal, the intense beam in zero-order direction was blocked by a beam-stopper (BS). To improve signal-to-noise ratio, we used a combination of a chopper wheel in the pump beam path and a boxcar averager in ``toggle''-mode. We additionally corrected the signal for laser fluctuations. The transients shown in Fig. 3 are averaged over 2000 laser pulses per delay step.

**Results and Discussion**

Figure 2-c shows the interference pattern between ± 1 order diffraction created from the 800 nm pulses with tilted pulse fronts in the PM, imaged onto a CCD camera. The image clearly shows a modulation in intensity along the horizontal axis. We estimate a groove density of ~100 lines/mm (~10 µm pitch), which corresponds to an angle of incidence of $\theta_{pu} = 2.29°$ for the ± first order diffraction in agreement with our angle of incidence determinations. In order to test the effectiveness of the pulse-front tilt approach, we performed transient grating measurements at room temperature.

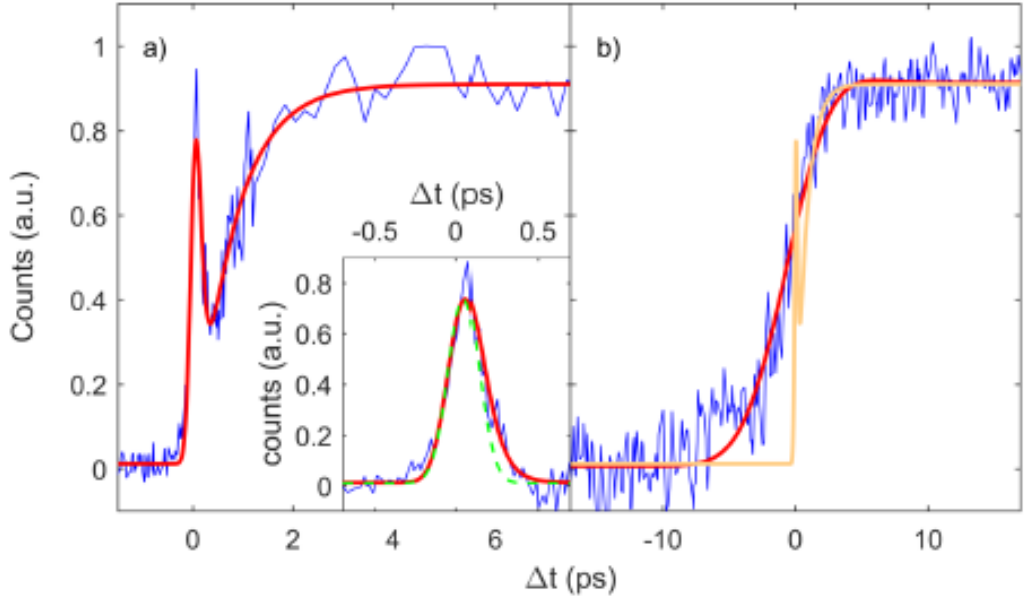

*Figure 3: Comparison of transients with matched (a) and unmatched (b) pulse fronts: In the first case the spectrum shows sub-ps features demonstrating that the time resolution of the tilted pulse-front setup is short enough to resolve ultrafast dynamics. In the second case we estimate a Gaussian response function of σ ~ 2.8 ps dominating the spectral shape. For the sake of comparison, we include in the plot, in orange, the fit function of Fig. 3-a. In the inset we highlight the ultrafast component of spectrum showed in (a) subtracted by the longer time scale feature. We used a three states rate equation model convoluted with a gaussian-like instrumental function to fit the data (in red). We revealed an ultrafast component with a characteristic time of 80 fs and estimated a response function (green dashed line in the figure) with σ of 90 fs.*

By rotation of BG and thereby either directing the 1st or 0th diffraction order onto the PM, the effect of pulse front matching can be directly assessed. Fig. 3 shows a comparison between transients with and without matched pulse fronts.

In Fig. 3-a we show a transient obtained with matched pulse fronts. At zero pump-probe delay, we observe a rapid turn on of transient diffraction intensity followed by a decay on the sub-ps timescale. The transient further evolves in form of a second, slower signal increase reaching a plateau at ≈ 3 ps. To quantify the time-dependence of our results, we fit them with a Trust-Region-Reflective Least Squares Algorithm using the following phenomenological function

$$g(t) = G(t) \otimes H(t)\left(Ae^{-\frac{t}{\tau_1}} + Be^{+\frac{t}{\tau_2}} + C\right), \tag{5}$$

in which is included the convolution by the instrumental response function approximated by a Gaussian G(t). H(t) is the Heaviside step function, $\tau_1$ and $\tau_2$ exponential time constants, and A, B, and C amplitudes. The fitted time constants are $\tau_1$ = (80 ± 40) fs and $\tau_2$ = (740 ± 90) fs. The fitted width of the instrument response function is σ = (90 ± 10) fs.

We use the fit results to subtract the signal evolution on the time scale of several hundred fs from the sub 100 fs dynamics, which is shown in the inset of Fig. 3-a. Comparison with the pure instrument response function, which is shown as a dashed green line, clearly shows a broadening of the feature towards positive pump-probe delays due to ultrafast dynamics in the sample.

For comparison, we show in Fig. 3-b a transient obtained with the 0$^{th}$ diffraction order of the BG i.e. unmatched pulse fronts. It is important to notice the different time scales of the two plots in Fig. 3-a and Fig. 3-b. To allow a quick comparison between the two spectra, we report the fit obtained for the tilted-pulse front case in orange in Fig. 3-b. It is evident that the unmatched pulse fronts prevent observation of any sub-ps dynamics. The signal shape is now clearly dominated by the instrument response function. In order to estimate the time resolution of the unmatched pulse front case, we applied the same fit keeping the time constants fixed at the values obtained for the tilted pulse-front case of Fig. 3-a. It results in a width of the instrument response function of σ = (2.8 ± 0.2) ps, in agreement with our earlier results (1). The fit is shown as a red line in Fig. 3-b.

It is not the goal of this work to clarify the microscopic origin of the spectral shape shown in Fig. 3-a. Nevertheless, we can qualitatively compare our results with the ones obtained with ultrafast transient holography experiments available in the literature (30–33). The technical conditions among our work and the ones cited here are different in many respects, including the optical setups, the crystal orientations, the thickness of the samples, the fluences, the laser wavelength and the substrates. It is not surprising that there is not a quantitative agreement on the measured time scales in play. Nonetheless, it is very clear that the shapes of the signals are very similar: all the experiments, as in our case, present a sharp peak at time zero (in our case we are able to measure also its decay profile) and a slower rising feature in the sub-ps regime. The first ultrafast feature is addressed to the formation of bipolaronic charge-transfer excitons, while the sub-ps dynamic should be related to the IMT process, possibly mediated by electron-phonon coupling.

**Conclusion**

We demonstrate in this contribution a versatile optical setup for investigating ultrafast dynamics on surfaces. The setup allows for the highest possible transient signal levels by probing the surface in grazing incidence, a crucial condition for experiments with comparably weak HHG sources. Its strongly noncollinear geometry allows for optimal time resolution while keeping the highest possible line density. In addition, the beams are mostly easily accessible outside the vacuum system for future EUV or SXR probing. The pulse front mismatch resulting from the noncollinear geometry is compensated by tilting the pulse fronts of the optical pump pulses in an easy-to-align setup. We demonstrate matching of the pulse fronts of pump and probe pulses with all-optical experiments on the insulator-to-metal transition in a thin film of VO$_2$.

**Acknowledgements**

This work was supported by the AMOS program within the Chemical Sciences, Geosciences and Biosciences Division of the Office of Basic Energy Sciences, Office of Science, U.S. Department of Energy. MG is funded by a Lichtenberg Professorship from the Volkswagen foundation. We thank Emily Sistrunk-Link and Jakob Grilj for fruitful discussions.